\documentclass[12pt]{article}
\usepackage{geometry}
\usepackage{amsfonts}
\usepackage{amsmath}
\usepackage{amssymb}
\usepackage{amsthm}
\usepackage{mathrsfs}
\usepackage{bbm}
\usepackage{cancel}
\usepackage{graphicx}
\usepackage{hyperref}
\usepackage[usenames,dvipsnames]{color}

\newcommand{\st}[1]{\text{\tiny \rm #1}}

\def\nb{$N$-body problem~}
\def\nbn{$N$-body problem}

\def\bq{\begin{equation}}
\def\ee{\end{equation}}
\def\m{$I_\st{cm}$ }

\def\rms{$\ell_\st{rms}$ }

\def\b{$S_\st{B}$ }
\def\bn{$S_\st{B}$}

\hypersetup{
colorlinks=true,         
linkcolor=blue,          
citecolor=red,        
urlcolor=Violet            
}

\hypersetup{
colorlinks=true,         
linkcolor=blue,          
citecolor=red,        
urlcolor=Violet            
}

\begin{document}

\title{\vspace{-1.5cm}\bf Arrows of Time in Unconfined Systems}

\author{{Julian Barbour,$^{1,2}$
{\rm }
}\vspace{12pt} \\
\it \small $^1$College Farm, South Newington, Banbury, Oxon, OX15 4JG UK,\\
\it \small $^2$Visiting Professor in Physics at the University of Oxford, UK.\\
}
\date{}

\maketitle

\begin{abstract}

Entropy and the second law of thermodynamcs were discovered 
through study of the behaviour of gases in confined spaces. The related 
techniques developed in the kinetic theory of gases have failed to resolve the 
apparent conflict between the time-reversal symmetry of all known laws of 
nature and the existence of arrows of time that at all times and everywhere in the universe all point in the same direction. I will argue that the failure may due to unconscious application to the universe of the conceptual framework developed for confined systems. If, as seems plausible, the universe is an unconfined system, new concepts are needed.

\end{abstract}

\maketitle

\section{Introduction\label{int}}

This paper is to appear in the proceedings of the Time in Physics conference held at the ETH Zurich from 7th to 11th September 2015. I do not attempt to cover all the ground of my talk, which can be viewed online \cite{video}, or the material in \cite{arrows, prl, entaxy} on which my talk was based. Instead, taking an historical perspective, I want to indicate why I think the traditional understanding of entropy needs to be modified if it is to be applied to the universe. The main reason is that thermodynamics and its interpretation by statistical mechanics were developed for confined systems whereas the universe appears to be unconfined. This, I believe, has far-reaching implications for all questions relating to the various arrows of time.

Simple examples explain what I mean by confined and unconfined systems. In the ideal-gas model, many particles move inertially apart from short-range elastic interactions. They are confined to a box at rest in an inertial frame and bounce elastically off its walls. That's a confined system. The same particles without box is an unconfined system. Pointlike particles that interact solely through Newtonian gravity can model an unconfined `island universe', but the ideal gas will already indicate the need for new concepts. Proper application of entropic ideas to the universe will surely need inclusion of gravity. My collaborators present ideas about that in \cite{entaxy} and about the quantum mechanics of unconfined systems in \cite{arrows}, Sec.~4.

\section{Properties of Confined and Unconfined Systems} 

My survey of the arrow-of-time literature failed to identify any study that highlights the distinction between confined and unconfined systems. True, the universe's expansion, aided by gravity, is often mooted (see, e.g., \cite{davies, zeh}) as the `master arrow' for the other arrows, but one finds little suggestion that the very concept of entropy needs reexamination in unconfined systems. The unconfined ideal gas shows that it does.

For this the heterogeneity of its degrees of freedom (dofs) is important: $N$ particles in Euclidean space have $3N$ Cartesian coordinates. Three locate the centre of mass, three define orientation and one overall size. If ${\bf r}_a^\st{\,cm}$ is the centre-of-mass position of particle $a$, the centre-of-mass moment of inertia (half the trace of the inertia tensor):  
\bq
I_\st{cm}=\sum_{a=1}^Nm_a\,{\bf r}_a^\st{\,cm}\cdot{\bf r}_a^\st{\,cm}\equiv{1\over m_\st{tot}}\sum_{a<b}m_am_b\,r_{ab}^2,~~ m_\st{tot}=\sum_a\,m_a,
\ee
or its square root (divided by the total mass), which is the root-mean-square length
\bq
\ell_\st{rms}=\sqrt{{\sum_{a<b}m_am_b\,r_{ab}^2\over m_\st{tot}^2}},
\ee
measures the size. The remaining dofs describe the \emph{shape} of the instantaneous configuration. This paper is about the different behaviours of the scale and shape dofs in confined and unconfined systems. It is revealing that even when general-relativistic cosmological models of unconfined universes have been considered an important consequence of the shape/scale difference has hardly ever been noted, as I shall explain. In books that do not consider cosmology, I have not once seen attention drawn to the shape/scale difference. As in many dynamical-systems studies, virtually all authors use Lagrange's generalized coordinates, which are simply denoted $q_1,\dots,q_n$. This hides all trace of the shape/scale difference.

\section{The Effect of Confinement\label{ec}}

Suppose that at $t_0$ all the particles of the ideal-gas model have random velocities and are in a small cloud in the centre of the box. The particles will spread out. Their elastic collisions with each other and the box walls will soon establish thermal equilibrium. Coarse graining will permit definition of a Boltzmann entropy $S_\st{B}$. It will be low at $t_0$ and then grow to a more or less stable maximum value. Since the system is perfectly isolated (affected by no external forces), it will be subject to Poincar\'e recurrence. Mostly, $S_\st{B}$ will exhibit very small fluctuations about its maximum with rare deep fluctuations. If at any time $t>t_0$ all the velocities are exactly reversed, the system will retrace its evolution back to the state of low $S_\st{B}$ at $t_0$, after which $S_\st{B}$ will rise again and embark on the typical Poincar{\'e}-recurrence behaviour of the forward time direction. The complete $S_\st{B}(t)$ curve, like the flanks of the entropy dips within it, will be qualitatively symmetric with respect to the direction of time. Note that the purely dynamical moment of inertia $I_\st{cm}$ (or $\ell_\st{rms}$) behaves just like the statistical $S_\st{B}(t)$. 
 
In this scenario, \emph{three} factors create the $S_\st{B}$ curve: the initial spreading of the particles into more phase-space cells; the interparticle collisions;  the particle--box collisions. Without the box, the interparticle collisions would soon cease and the sole cause of $S_\st{B}$ growth would be the growth of $\ell_\st{rms}$.

\section{Conceptual Inertia}

The discovery and statistical interpretation of entropy by Carnot, Lord Kelvin, Clausius, Maxwell, Boltzmann and Gibbs was a huge triumph. But for study of an expanding universe, are their concepts still appropriate? Consider \emph{thermodynamic systems}. Crucially important are their properties \cite{fermi}, first among them the volume $V$ of the studied substance. Without confinement, $V$ is not defined. Nearly as important are the pressure $p$ and temperature $T$. Like $V$, these \emph{state functions} cannot be determined without confinement, either by man or by nature through crystalline `self-containment'. Also essential, as in a Carnot cycle, is the possibility of reversible change of one state function while keeping the others fixed.

Confinement is crucial: without it \emph{equilibrium cannot be established} and state functions determined. Moreover, thermodynamic entropy $S$ can only be defined relative to a \emph{reference state}, from which the system must be carried reversibly through equilibrium states to the current state. In classical thermodynamics, $S$ is therefore defined only up to an additive constant. Quantum mechanics did eliminate this ambiguity through the 3rd law of thermodynamics, which defines the reference state as the one at the absolute zero of temperature at which there is only a single ground state (or so few that the logarithm in the definition of $S$ makes their contribution negligible). However, quantum mechanics did not lessen the need for confinement. Definite quantum states only exist if their wave functions satisfy appropraite conditions at some physical boundary, as happens, e.g., with phonon states in a crystal. In summary, statements in thermodynamics can only be reliably made if the system is confined and in equilibrium.\footnote{\,Fermi's definition of the entropy \cite{fermi} of out-of-equilibrium systems is illuminating. They must consist of subsystems each in equilibrium and separated by heat-insulating walls.} 

Confinement in thermodynamics has a counterpart in statistical mechanics, which requires \emph{normalizable} probability functions: the Liouville measure of the space of accessible states must be bounded.\footnote{This is also the most important condition required for Poincar{\'e}'s recurrence theorem to hold.} On this basis, Gibbs \cite{gibbs} required confinement of the system to a finite region of space\,\footnote{\,Gibbs noted that this restriction has a counterpart in thermodynamics, in which ``there is no thermodynamic equilibrium of a (finite) mass of gas in an infinite space''.} and a bound on the momenta (thereby ruling out systems with $1/r^2$ forces and, with them, self-gravitating systems).

I have recalled this history in order to consider whether the methods developed for confined systems can be sensibly applied to the universe. As regards the concepts and methods of thermodynamics, it is obvious that human control of state functions and enforcement of equilibrium are out of the question. However, `self-containment' can be said to occur in two situations. 

The first relies on the notion of a comoving volume in an expanding homogeneous unverse filled with either blackbody radiation or a non-relativistic ideal gas in local thermodynamic equilibrium. Describing work of Tolman, Davies \cite{davies}, writes

\begin{quote}\small

In the real universe there are no comoving \emph{bounded} volumes in which we 
may imagine the radiation to be confined. Instead, we may imagine an invisible 
transparent 2-surface delineating the volume, with photons continually 
crossing to escape from the inside. However, if the space is homogeneously 
filled with radiation, photons will enter the volume from the outside at the 
same rate, so the average number of photons in any such volume is constant.
\end{quote}\normalsize 

Tolman showed that in such comoving volumes the entropies of both blackbody radiation and an ideal non-relativistic gas in thermal equilibrium remain constant. Provided the departure from homogeneity remains small and does not significantly perturb the microwave background, it is possible to give a sensible estimate of the entropy and its growth within our Hubble radius.

The other example of self-containment is associated with horizons: most confidently with the event horizon of black holes and rather less with the particle horizon in de Sitter space and the Rindler-wedge horizon of a uniformly accelerated observer in Minkowski space. The entropic interpretation associated with horizons in these cases relies to some extent on information-type arguments: the entropy is said to represent an observer's ignorance of what is on the unobservable `other side' of the horizon. Although few theoreticians doubt the existence of a deep connection between gravity, thermodynamics and entropy, it may be noted that the beautiful proofs which lead to this confidence rely in part on assumptions that can be questioned. In particular, it is often assumed that space is asymptotically flat. Also the black hole event horizon is not so much inpenetrable as semipermiable (matter can fall into the black hole) and requires a subtle definition involving the state of the complete universe long after the collapse of matter that leads to the formation of the black hole.

The universally recognized problem is a general definition of gravitational entropy. This is widely attributed to the breakdown of homogeneity at the end of the era well described by FLRW cosmologies. I will suggest that a much more serious problem is the very definition of entropy in a system that can expand freely. Since the universe is manifestly far from equilibrium once it becomes inhomogeneous, thermodynamic concepts which rely on equilibrium will not help. We must see how far we can get with concepts taken from statistical mechanics.

\section{Unconfined Systems}

To this end, let us now consider unconfined systems, starting with the very simplest: two particles moving inertially. This system exhibits a feature that will occupy a central position in my discussion, both conceptually and literally: in every solution there is a unique instant that divides every solution into two halves. This is the instant at which the two particles are closer to each other than at any other time. 

This is a very trivial system, but it already exhibits the feature I want to highlight. We get a more illuminating example if we add Newtonian gravity, for which the two-body solutions are of three kinds depending on the total centre-of-mass energy $E_\st{cm}$: elliptical ($E_\st{cm}<0$), parabolic ($E_\st{cm}=0$) and hyperbolic ($E_\st{cm}>0$) motion of each particle about the common centre of mass. The elliptical case is periodic and quite different to the other two but does have successive points of closest approach that each divides the current orbit in half. In the other two cases, there is always a unique point of closest approach. Even the case of collision can be regularized by a bounce, which maintains the rule.

The $N$-body problem, $N\ge 3$, is much more interesting. It hardly ever enters university dynamics courses, which pass directly from two-body problems to rigid-body theory and then to Lagrangian and Hamiltonian theory. This may explain why a fact with a possibly deep connection with the second law of thermodynamics has escaped attention. I recall first that a potential $V({\bf r}_a)$ is homogeneous of degree $k$ if, for $\alpha>0$,  $V(\alpha{\bf r}_a)=\alpha^kV({\bf r}_a)$. For any such potential, Newton's 2nd law leads to the relation
\bq
\ddot I_\st{cm}=2E_\st{cm}-(k+2)V.\label{lj}
\ee
For the Newton potential $V_\st{New}, k=-1$. Thus, in the $N$-body problem, $\ddot I_\st{cm}=2E-V_\st{New}$. In addition, $V_\st{New}$ is negative definite, so if $E_\st{cm}\ge 0$
\bq
\ddot I_\st{cm}>0.\label{pos}
\ee 
This means that the graph of $I_\st{cm}$ as a function of the time $t$ is concave upwards and tends to infinity in both time directions. This fact, first discovered for 3-body motions in 1772 by Lagrange and later generalized to the \nb by Jacobi, was the first qualitative discovery made in dynamics and played an important role in the history of dynamics because it showed that the \nb with $E_\st{cm}\ge 0$ is unstable: at least one particle must escape to infinity. This then raised the question of whether the solar system, for which $E_\st{cm}<0$, is stable, the study of which led to Poincar{\'e}'s discovery of chaos. Another important consequence of (\ref{pos}) is the monotonicity of $\dot I_\st{cm}$:
\bq
{1\over 2}\dot I_\st{cm}=D. \label{dil}
\ee
The monotonic quantitiy (\ref{dil}), which by its close analogy with angular momentum may be called the \emph{dilational momentum}, is a Lyapunov variable; its existence immediately shows that there can be no periodic motions or Poincar\'e recurrence in the \nb with non-negative energy. For inertial motion, for which $V={\rm{const}}, k=0,$ so in this case too (\ref{pos}) holds and \m has a unique minimum. 

In \cite{entaxy}, my collaborators and I coined the expression \emph{Janus point} for the minimum of $I_\st{cm}$ and \emph{Janus-point systems} for unconfined dynamical systems for which \emph{every} solution divides into two (qualitatively similar) halves at a unique central point. Moreover, as pointed out in \cite{arrows, prl, entaxy}, the evolution in either direction away from the Janus point $J$ is \emph{time-asymmetric} even though the governing equation is time-reversal symmetric. This can be seen very easily in purely inertial motion, in which the position vector of each particle satisfies ${\bf r}_a(t)={\bf r}_a^0+{\bf v}_at$, where ${\bf r}_a^0$ is the initial position and ${\bf v}_a$ the (constant) velocity. With the passage of time (in either direction $t\rightarrow\pm\infty$), the contribution of the velocity term must become dominant. Moreover, because the particles with greater velocities get ever further from the slower particles, the rate of separation $\dot r_{ab}$ of any two particles $a$ and $b$ tends to become ever more closely proportional to their mutual separation $r_{ab}$:~$\dot r_{ab}\propto r_{ab}$. This Hubble-type expansion will occur not only in inertial motion but also for an ideal gas if the confining box is suddenly removed.

The time asymmetry either side of the minimal $I_\st{cm}$ at $J$ is therefore manifested in the ever greater tendency to Hubble-type expansion away from $J$. Moreover, the system is always in its \emph{most disordered state} around $J$. In the \nbn, the effect is much more striking because bound clusters are formed and move away from each other in Hubble-type expansion. This causes growth (between bounds that grow as $t\rightarrow\pm\infty$) of a scale-invariant quantity called \emph{complexity} in \cite{arrows, prl, entaxy}.

There is a deep reason for the time-asymmetric behaviour: Liouville's theorem. In accordance with what I said about degrees of freedom, the total phase-space volume is divided into parts: an orientational part (which we can ignore), a shape part and the scale part. At $J$, the scale variable $\ell_\st{rms}$ takes its minimal value and increases monotonically in both directions away from $J$. Given a Gibbs ensemble of identical systems at $J$, the phase-space scale part must increase as $t\rightarrow\pm\infty$. This means that the shape part must decrease: dynamical attractors must act on the shape degrees of freedom.\footnote{\,That growth of the scale part of phase space must reduce the part corresponding to the remaining degrees of freedom was noted in connection with inflation in \cite{sloan}.} In \cite{arrows, prl, entaxy}, it is shown that arrows of structure formation must emerge through this effect. Whether all known arrows emerge in this way remains to be seen. If they do, expansion of the universe will indeed be the master arrow responsible for them.

In this connection, it is important that overall scale cannot be observed for observers within a universe. Observed facts are ratios.\footnote{\,This was the main motivation for the development of \emph{shape dynamics} \cite{GGK,tutorial}.} 
One reason we say the universe is expanding is that the ratio of the intergalactic separations to the galactic diameters is growing. Moreover, expansion of the universe was first deduced from red shifts, which are ratios of wavelengths. Thus, as we address the problem of defining an entropy-type concept for the universe, we must take into account two facts: 1) only shape variables, which are dimensionless ratios, can be accessed by observers within the universe; 2) in an expanding universe, the shape variables are subject to attractors.

\section{Implications for the Definition of Entropy}

As noted at the end of Sec.~\ref{ec}, if all the particles of an ideal gas are situated at $t_0$ in a small region within a much larger box three factors contribute to the $t>t_0$ behaviour of $S_\st{B}$: the initial more or less free growth $\ell_\st{rms}$ with some  interparticle collisions; the particle collisions with the box walls once $\ell_\st{rms}$ is large enough; thereafter regular interparticle and particle--wall collisions with essentially constant $\ell_\st{rms}$.

There is a common intuition that entropy increase corresponds to growth of disorder. Random motion of particles in a confined region seems much more disordered than free Hubble-type expansion. In the previous paragraph's scenario, disorder-increasing interparticle and particle--box collisions rapidly erase the initial expansion's disorder-decreasing effect. However, in the absence of a box the latter rapidly becomes the dominant effect. This simple observation suggests that entropic concepts need reconsideration if thay are to be applied to a freely expanding universe.

This can be seen especially clearly if we include gravity and model the universe by the \nb with $E_\textrm{cm}=0$. As we have seen, it is an immediate consequence of Liouville's theorem that the shape of the system is attracted to ever smaller regions of the system's space of possible shapes (shape space {\sf S}) with increasing distance from the Janus point $J$. Intuitively, this is anti-entropic behaviour. Indeed, in \cite{entaxy} my collaborators and I use the scale-invariant complexity mentioned earlier as a state function to define a Boltzmann-type count of microstates we call \emph{entaxy} (to avoid confusion with the entropy concept that can be meaningfully used for confined systems). We argue that entaxy, not entropy, must be used to characterize the typicality of the universe's state. What is more, the entaxy always has it greatest value near $J$ and \emph{decreases} in both directions away from it. At the same time, the universe becomes more structured because bound subsystems form and separate from each other in Hubble-type expansion.

Thus, \emph{as the universe evolves in both directions away from J, its complexity increases while its entaxy decreases.} There is nothing mysterious about this inversion of normal entropic behaviour. It is due to the difference, enhanced by gravity, between confined and unconfined systems. We also point out in \cite{entaxy} that the subsystems which gravity creates become more or less `self-confined'. As I noted earlier, this is the \emph{sine qua non} for application of Gibbs-type statistical-mechanical arguments based on conventional entropic notions. In fact, we are able to show that the subsystems form with some given Boltzmann entropy \bn, which then \emph{increases}.  Moreover, the overwhelming majority of these subsystem entropies all increase in the same direction as the universe's entaxy decreases. This shows how local entropy increase -- the tendency of a confined system's state to become less special -- is compatible with the simultaneous tendency of the universe to become more special.

This also casts light on our experienced direction of time. Boltzmann argued that it is aligned with the direction of increasing entropy. The apparent conflict with the growth of records and structure we see around us is widely said to be perfectly compatible with the 2nd law: a decrease of $S_\st{B}$ here is more than compensated by an increase elsewhere. This is often stated without proof. When one is given, it often invokes refrigerators, in which the cooling is more than offset by the heating of the environment. But if this is to be quantified, the environment must be confined, since otherwise its increase in $T$ and $S_\st B$ cannot be determined. In the absence of physical insulating walls, we are back to the problem of defining the universe's entropy.\footnote{\,Planck's well-known statement of the second law shows how essential it is to have complete control over the environment: ``It is impossible to construct an engine which will work in a complete cycle and produce no effect except the raising of a weight and cooling of a heat reservoir.''} The mismatch between the universe's increasing structure and the entropic arrow is resolved in \cite{entaxy}. Entaxy determines the \emph{master arrow}. In a self-gravitating universe it creates more or less stably bound subsystems. In turn, these are born with a certain \b that in the overwhelming majority of cases then increases in the same direction as the master arrow which gave birth to them. Moreover, the Janus-point structure (and with it the oppositely pointing master arrows) is a dynamical necessity. It is not imposed by a special selection principle. It merely requires a non-negative energy and \emph{an unconfined system}.

In discussing `conceptual inertia', I noted that collisions tend to increase disorder but growth of \rms has the opposite effect. Could it be that the almost exclusive concentration on confined systems in statistical mechanics has allowed this difference to escape notice? I have not studied the literature exhaustively, but I found few discussions of the entropy of a freely expanding gas.

Gibbs, as we saw, ruled out systems in infinite space in order to avoid unnormalizable probability functions. However, Tolman \cite{tolman}, having noted that in confined systems entropy will increase to an equibrated maximum, then continued ``in the case of unconfined gases \dots a final state of infinite dilution and complete dissociation into atoms would be one of maximum entropy''. Davies \cite{davies}, p.~33, discussing the explosive escape of gas from a cylinder says ``the second law becomes an expression of the principle that a gas will explode into a vacuum, but will never spontaneously implode into a smaller volume''. Two comments can be made here. First, the gas under consideration forms a \emph{subsystem} of the universe; it does not serve as a model of the whole universe, in which (for a given choice of the nominal time direction) spontaneous implosion (followed by explosion) does occur. Second, Davies does not say explicitly that the entropy of the exploding gas increases, only that, in being irreversible, the process is an expression of the second law. Finally, discussing the inertial model discussed here and in \cite{entaxy} in the recent \cite{ns}, Carroll and Guth say the model exhibits a ``two-headed arrow of time'' in which entropy increases in both limits $t\rightarrow\pm\infty$ (see also \cite{snews}). 

That Janus-point solutions exhibit oppositely directed arrows of time can hardly be doubted, but whether one can say entropy increases in the direction of the arrows seems very questionable. I have already noted that traditional thermodynamics of the universe cannot exist because the universe is not a thermodynamic system whose state can be changed and measured. Application of conventional statistical mechanics to universes that can expand is also highly problematic because of the problem pointed out by Gibbs: probability distributions are only meaningful if they can be normalized, which means that they must be defined on a space with a bounded measure. 

At this point I will stop. My main point -- the need to think about the entropy and statistics of universes differently -- has been made. I will only say that the greatest difficulty to which I have drawn attention, the unbounded phase space of an expanding universe, may suggest \cite{arrows, prl, entaxy} its solution. For Liouville's theorem directs us to the attractor-induced arrows on the space {\sf S} of possible shapes of the universe, and {\sf S} is obtained by quotienting the Newtonian configuration space by translations, rotations and dilatations. Due to these last, the resulting space is \emph{compact}, so that one can define on it a bounded measure. As explained in \cite{video, arrows, prl, entaxy}, this meets Gibbs' requirement for meaningful definition of probability distributions and opens up the possibility of creating a theory of the statistics of universes.

\emph{Acknowledgement.} My thanks to Tim Koslowski and Flavio Mercati for the stimulating and fruitful collaboration that led to \cite{arrows, prl, entaxy}.

\end{document}